\documentclass[twocolumn,prl,showpacs,superscriptaddress]{revtex4}
\usepackage{graphicx} 
\usepackage{amsmath} 
\usepackage{amssymb}
\usepackage{amsbsy}
\usepackage[ansinew]{inputenc} 

\begin{document}
\title{Local Moment Formation and Kondo Effect in Defective Graphene}
\author{M. A. Cazalilla}
\affiliation{Centro de F\'isica de Materiales CSIC-UPV/EHU
and Donostia International Physics Center (DIPC). Paseo Manuel
de Lardizabal,  E-20018 San Sebastian, Spain}
\affiliation{Graphene Research Centre
National University of Singapore,
6 Science Drive 2,
Singapore 117546.}

\author{A. Iucci}
\affiliation{Instituto de F\'isica de La Plata (IFLP) - CONICET and Departamento de F\'isica, 
Universidad Nacional de La Plata, cc 67, 1900 La Plata, Argentina}
\author{F. Guinea}
 \affiliation{Instituto de Ciencia de Materiales de Madrid (ICMM). CSIC,
 Cantoblanco. E-28049 Madrid. Spain.}
 \author{A. H. Castro Neto}
\affiliation{Graphene Research Centre
National University of Singapore,
6 Science Drive 2,
Singapore 117546.}

\begin{abstract}
 We study the local moment  formation and the Kondo effect at single-atom vacancies in Graphene. 
 We develop a model accounting for  the vacancy reconstruction as well as non-planarity effects induced by strain and/or 
 temperature. Thus, we find  that  the dangling $\sigma$ orbital localized at the vacancy is allowed to strongly hybridize
 with the $\pi$-band  since the scattering with the vacancy turns the hybridization into 
 singular function of the energy ($\sim [|\epsilon| \ln^2 \epsilon/D]^{-1}$, $D\sim$ the bandwidth). 
 This leads to several new types of  impurity phases, which  control the magnitude of  the vacancy magnetic 
 moment  and the possibility of Kondo effect  depending on the strength of the local  Coulomb interactions, the 
 Hund's rule coupling,  the doping level, and the  degree of particle-symmetry breaking.  
 \end{abstract}
\date{\today} \maketitle

%

\begin{figure}[b]
\includegraphics[scale=0.3]{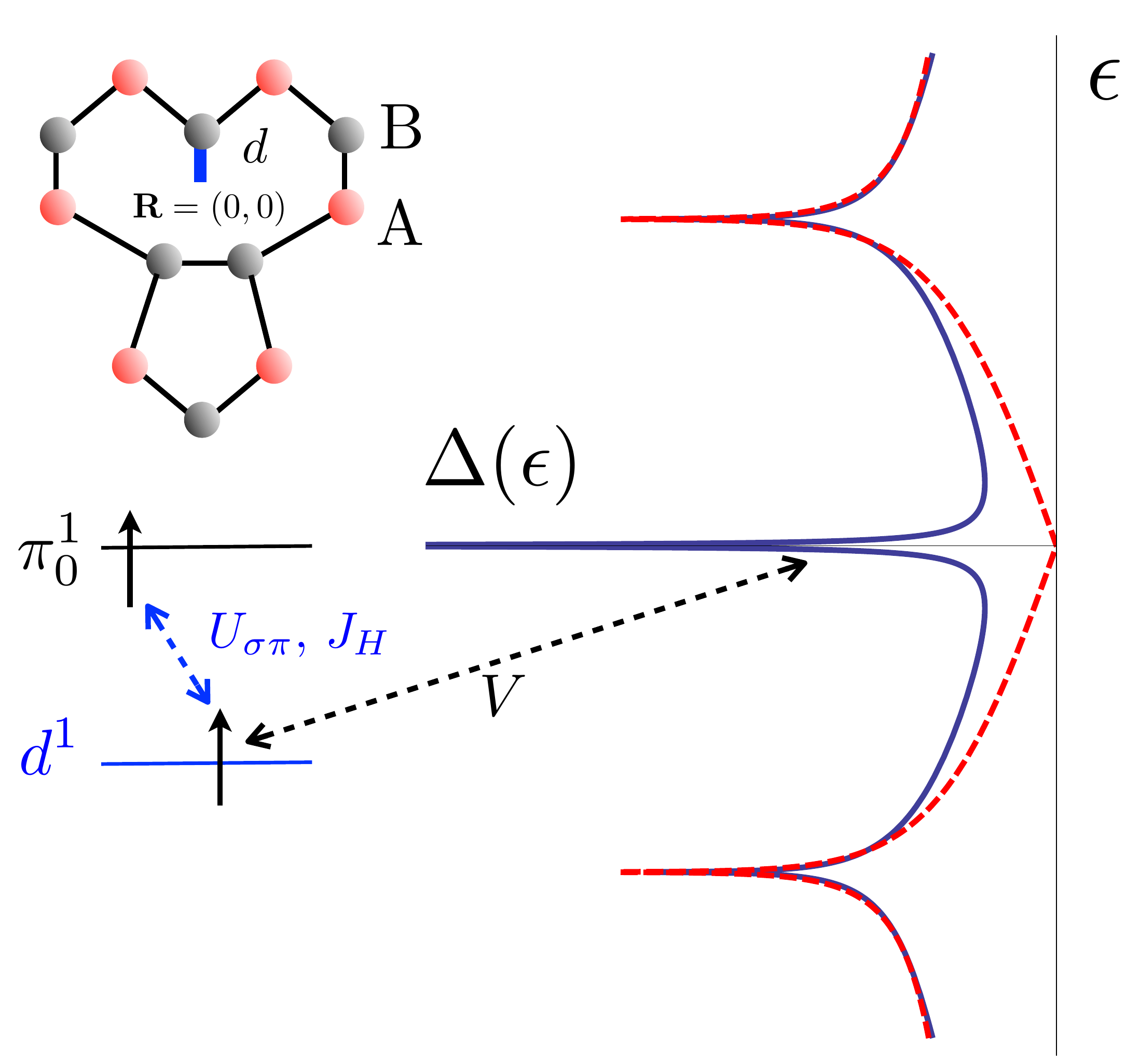}
\caption{Two level Anderson model describing a reconstructed single-atom 
vacancy in rippled graphene; $d$ is an orbital
that is mainly localized at a dangling $\sigma$ orbital of the apical atom; 
$\pi_0$ is the vacancy-induced localized state. Rippling implies that the
$d$-orbital can hybridize with the $\pi$ conduction band via sp$^2$-sp$^3$ hybridization, 
$V$. The hybridization, $\Delta(\epsilon)$, is energy  ($\epsilon$) dependent and enhanced 
by the strong scattering  potential at the vacancy. $U_{\sigma\pi}$ and $J_H$
describe the  Coulomb interaction between the two localized levels. See Eq.~\eqref{eq:am}.}
\label{fig:fig1}
\end{figure}
  Experimental evidence of  magnetic  local moments (MLM) in irradiated graphene
at very low temperatures  has been recently reported  in~\cite{Grigorieva,Kawakami}.  
In another experiment using irradiated samples~\cite{Fuhrer},  the Kondo effect with surprisingly
high Kondo temperatures ($T_K\sim 10$-$100$ K, obtained by fitting resistivity data to conventional spin-$1/2$ Kondo 
laws~\cite{Hewson}) was observed.   \emph{A priori}, the latter observation  appears to be  at odds with 
the  observations of~\cite{Grigorieva,Kawakami}. For instance, at the  temperatures of 
 the Manchester group experiment  ($T=2$K)~\cite{Grigorieva},   the measured spin-$1/2$ 
 MLM of the vacancies created by irradiation should be  quenched by the Kondo effect observed 
in~\cite{Fuhrer}. Furthermore, the authors of~\cite{Fuhrer} reported a
 $T_K$ with a rather weak dependence on  gate voltage. However, 
 theories of the Kondo effect in graphene~\cite{Fradkin,SenguptaBaskaran,Vojta2010,Uchoa2011}   
predict a critical coupling below which MLM are stable at the Dirac point (DP) 
whilst  $T_K$ remains small around the DP~\cite{SenguptaBaskaran,Vojta2010}.

\begin{figure}[t]
\includegraphics[scale=0.3]{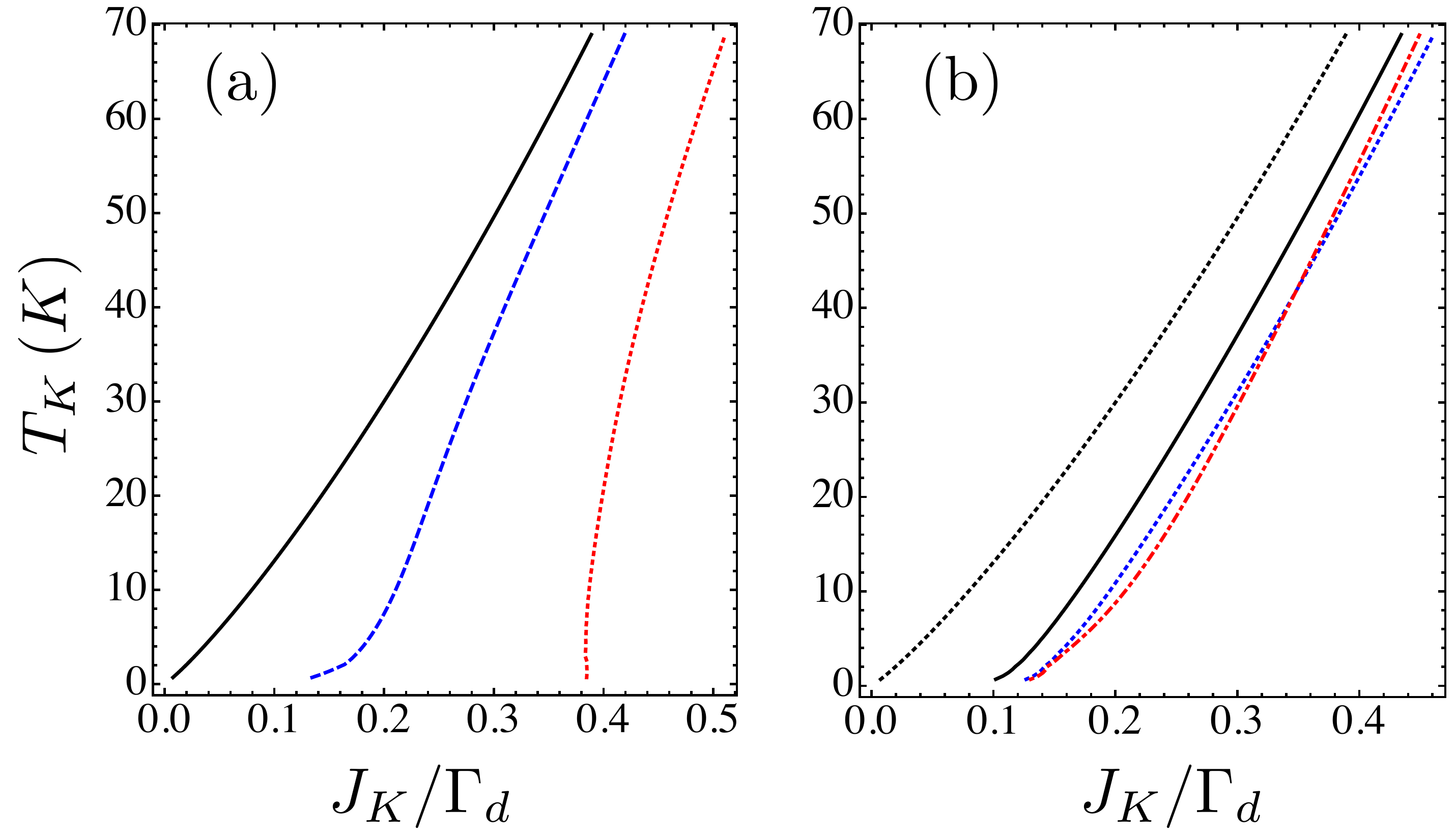}
\caption{For a vacancy with $S_v = \frac{1}{2}$, large $N$ mean-field Kondo 
temperature $T_K$ vs. $J_K/\Gamma_d$ ($J_K = 2 |V|^2/|\epsilon_d|$ and $\Gamma_d = \pi |V|^2 
\Delta(\epsilon_d)$,($\epsilon_d  = -0.75$ eV,
$t = 2.97$ eV and $U\to +\infty$).  Panel (a): $\mu = 0$ and $t^{\prime} = 0$ (continuous), 
$\mu = 7.4$ meV and $t^{\prime} = 0$ 
(dashed), and $\mu = 0$ and $t^{\prime} = 30$ meV. Panel (b): $t^{\prime} = 59$ meV 
and  $\mu = -30$ meV (continuous), $\mu =-22$ meV (dotted), $\mu = -37$ meV (dot-dashed). 
The $\mu, t^{\prime} =0$ result is also plotted for reference (dashed). Note 
that for $J_K/\Gamma_d\sim 1$ the vacancy enters the mixed-valence regime.}
\label{fig:fig2}
\end{figure}

 The  puzzle described above calls for re-visiting the magnetic properties of isolated
single atom vacancies in graphene.  Here we show that earlier 
theories~\cite{Fradkin,SenguptaBaskaran,Vojta2010,Uchoa2011} 
must be modified to  account for the strong scattering potential of the vacancy.
Indeed, the later is responsible for a dramatic change of the energy dependence of 
the hybridization with the $\pi$ band of a level localized at a reconstructed 
vacancy~\cite{Yazyev2007,Nanda2011,PalaciosYndurain}. Thus,  the energy $\epsilon$ dependence of 
the hybridization is changed to  $\sim [|\epsilon| \ln^2 (\epsilon/D)]^{-1}$ near the DP (cf. Fig.~\ref{fig:fig1}),
which is different  from the pseudogap behavior $\sim |\epsilon|^{r}$ ($r > 0$)  considered earlier~\cite{Fradkin,SenguptaBaskaran,Vojta2010,Uchoa2011}. Therefore,  it becomes possible for vacancy MLM to be 
quenched at  $T_K \sim 1-100$ K  (cf. Fig~\ref{fig:fig2}).

 The possibility of defect-induced magnetism in graphene has been widely discussed in the  
 literature (see e.g.~\cite{theory,Ochoa2011,Nanda2011,PalaciosYndurain}, and references therein),
 mainly using  various types of static mean-field theories.  However, a recent study~\cite{Ochoa2011}
 using dynamical mean-field theory concluded that vacancies should allow for the formation of MLM that
are \emph{ferromagnetically} coupled to the conduction band of graphene. Within Hartree-Fock, this happens
because of the strong local correlations~\cite{hubbardU}  and the large enhancement of the 
local density of states in the neighborhood of the vacancy~\cite{Pereira2006Nuno}. 
However, the analysis of Refs.~\cite{theory,Ochoa2011}  focuses only on the contribution to the
vacancy magnetic moment  from $\pi$-band electrons. 
\emph{Ab initio} calculations using the generalized gradient approximation (GGA) 
to density functional theory (DFT) ~\cite{Nanda2011,PalaciosYndurain}, can 
 account for the contribution of the dangling $\sigma$  orbitals  and are also able to describe the possible  
reconstructions of the vacancy~\cite{Zunger1978}. However, such studies predict different values 
for the vacancy MLM~\cite{Yazyev2007,Nanda2011,PalaciosYndurain}.

  Indeed, provided the  dangling $\sigma$ orbitals are not passivated,
the vacancy  reconstructs~\cite{Zunger1978,Nanda2011,dosSantos2011,PalaciosYndurain} 
following a Jahn-Teller distortion. Thus, a carbon pentagon with
a strong bond resulting from two of the dangling $\sigma$ orbitals (cf. Fig.~\ref{fig:fig1}) is formed.  The remaining
dangling $\sigma$ orbital is mainly localized at the apical atom (opposite to the 
reconstructed bond), and appears as a level  at  $\sim 1$ eV~\cite{Nanda2011,PalaciosYndurain}
below the DP (cf. Fig.~\ref{fig:fig1}). Double occupation of this orbital is strongly suppressed by a rather strong  
Coulomb repulsion $U \sim 10$ eV~\cite{hubbardU}, which leads to the formation of a MLM.  
However, in flat graphene,  symmetry forbids the hybridization of $\sigma$  and $\pi$ orbitals, meaning that
 the electron in the  dangling $\sigma$ orbital  cannot  hybridize with $\pi$ band and thus 
 exchange with conduction electrons. In absence of  other contributions to the  
 MLM~\cite{PalaciosYndurain},  this fact is hard to reconcile with the
observed  vacancy effects on charge~\cite{Fuhrer} and spin transport~\cite{Kawakami}.

 The above picture changes  if we recall that graphene is a membrane that  becomes  easily
rippled under the effect of strain and temperature~\cite{ripples}. Indeed, DFT-GGA calculations~\cite{dosSantos2011} 
indicate that, under rather small ($\approx 1$\%) isotropic compression, regions containing reconstructed vacancies  
do ripple~\cite{Pentagons}. 
Rippling  is achieved by means of sp$^2$-sp$^3$ hybridization of the carbon 
bonds~\cite{Pauling,CastroNetoGuinea}, which admixes the $\sigma$  with  p$_z$ orbitals at a given
atom, resulting in  new $\sigma$ orbitals pointing along directions that are no longer perpendicular to the $p_z$  orbital. 
Thus,  in a typical sp$^2$-sp$^3$ hybridized configuration, a carbon atom
can  stick out of the graphene plane. If the atom 
contains a dangling $\sigma$ orbital, as in the case of the apical atom of a reconstructed 
vacancy, the $\sigma$ orbital will be able hybridize with the $\pi$ band. 
In fact, the hybridization energy, $V_{\sigma\pi}$, can be substantial already for small 
deviations from planarity~\cite{Pauling,CastroNetoGuinea}:
$\tilde{V}_{\sigma\pi} = A \sqrt{\frac{1-A^2}{3}} (\epsilon_{s}-\epsilon_{p})$,
where $\sin \theta = A/\sqrt{A^2+2}$,  $\theta$ being the angle subtended by the dangling $\sigma$ orbital 
atom and the graphene plane, and $\epsilon_s-\epsilon_p = -8.31 eV$ is the atomic 
s-p splitting. Thus, in rippled graphene  $\pi$ electrons are allowed to hop on and off the dangling $\sigma$ orbital,
thus leading to an effective (anti-ferromagnetic, AF) exchange  with the conduction band.
However, this also means that  the MLM of the dangling $\sigma$ orbital
may be quenched by the Kondo effect. Indeed,  this possibility is not negligible, since 
the strong scattering potential  of the nearby vacancy  substantially 
modifies the local density of states~\cite{Pereira2006Nuno}.  

 Following the previous discussion, we introduce a model that  contains all the ingredients  described 
above~\cite{suppl} and   whose  Hamiltonian can be written as $H = H_{d} +  H_{\mathrm hyb} + H_{V+\pi}$, 
where~\footnote{The convention of  summing over repeated Greek indices ($\alpha=\uparrow,\downarrow$) 
will be used throughout.} 
\begin{align}
H_{d} &= \epsilon_d n_d + U n_{d\uparrow} n_{d\downarrow} + H_{d\pi},\\
H_{\mathrm{hyb}} &= V \left[ d^{\dag}_{\alpha} b^{\alpha}_0 + b^{\dag}_{\alpha} d^{\alpha} \right]\\
H_{V+\pi} &=  \varepsilon_0\,  a^{\dag}_{\alpha 0} a^{\alpha}_0 -t \sum_{\langle i,j\rangle} a^{\dag}_{i \alpha} b^{\alpha}_{j}  
+ \cdots
\end{align}

The term $H_d$ above describes the dangling $\sigma$ orbital with 
 $\epsilon_d \simeq -0.75$ eV measured from the DP~\cite{Zunger1978}, and $U\approx 10$ eV~\cite{hubbardU};
$n_{d\uparrow(\downarrow)} = d^{\dag}_{\uparrow(\downarrow)} d^{\uparrow(\downarrow)}$;
$H_{d\pi}$ (see below) describes the Coulomb interaction  between the localized dangling $\sigma$ orbital ($d$-orbital, in 
what follows) and the $\pi$ band electrons  below. 
$H_{\mathrm hyb}$ is the hybridization with the $\pi$-band, where
$V = \varphi_d(\boldsymbol{0}) \tilde{V}_{\sigma\pi}$ ($|\varphi_d(\boldsymbol{0})|^2\simeq 0.7$~\cite{Nanda2011,suppl}). In the last term , the limit $\varepsilon_0 \to +\infty$ must be taken in order to describe a single-atom vacancy on 
the A sublattice at $\mathrm{R} =  (0,0)$. The apical atom corresponds to the
B sublattice atom within the same lattice. The fermion operators $a^{\alpha}_i, b^{\alpha}_i$ ($a^{\dag}_{\alpha i}, b^{\dag}_{\alpha i}$)
destroy (create) $\pi$-band electrons of spin $\alpha$ on sites belonging to A and B sublattices. 

 To begin with, we retain only the nearest neighbor hopping ($t \simeq 2.8$ eV~\cite{RMP}) and 
consider  the chemical potential to be at the DP (i.e. $\mu = 0$). The effect of a nonzero $\mu$ 
 and  the next-nearest neighbor hopping  $ - t^{\prime} \sum_{\langle\langle i, j \rangle\rangle} \left[ a^{\dag}_{i \alpha} a^{\alpha}_{j} +  b^{\dag}_{i \alpha} b^{\alpha}_{j}  \right]$  will be discussed at the end. Thus, the spectrum of
$H_{\pi+V}$ contains a localized level (a zero-mode, ZM) pinned at the DP, which is
orthogonal  to a  continuum of waves scattering off the vacancy~\cite{Pereira2006Nuno}. 
The ZM wavefunction is not square normalizable since  
$|\varphi_0(\mathbf{R})|^2 \sim \frac{1}{|\mathbf{R}|^2}$, but its overlap with the $d$-orbital is the largest, 
which means that the dominant contribution to $H_{d\pi}$ is 
from the Coulomb interaction between the electrons in the ZM and the  $d$-orbital~\cite{suppl}, i.e.
$H_{d\pi} = U_{\sigma\pi} n_{d} n_0  - J_{H} \mathbf{S}_d \cdot \mathbf{S}_{0}$, where $n_{d} = 
n_{d\uparrow}+n_{d\downarrow}$, $n_0 = \pi^{\dag}_{\alpha 0} \pi^{\alpha}_0$, where $\pi^{\alpha}_0$ ($\pi^{\dag}_{\alpha 0}$)
destroys (creates) electrons in the ZM,  and $\mathbf{S}_{d} = d^{\dag}_{\alpha}\left(\frac{\boldsymbol{\sigma}}{2} \right)^{\alpha}_{\beta}
d^{\beta}$, and $\mathbf{S}_{0} = \pi^{\dag}_{0\alpha}\left(\frac{\boldsymbol{\sigma}}{2} \right)^{\alpha}_{\beta}
\pi_0^{\beta}$, where $\boldsymbol{\sigma} = (\sigma^x,\sigma^y,\sigma^z)$ are the Pauli matrices.
$J_H > 0$ is the Hund's coupling and $U_{\sigma\pi} > 0$ is the Coulomb repulsion between 
electrons in the ZM and the $d$ orbitals. The relative value of these parameters determines whether the
vacancy spin is a doublet or a triplet. However, the extended nature of the ZM makes it difficult to obtain accurate estimates for 
$J_H$ and $U_{\sigma\pi}$ (some estimates, obtained on dimensional grounds, are provided in~\cite{suppl}). Thus,  in absence of  accurate input, below we explore the possible magnetic
phases  for both the doublet and the triplet. 
 
Treating electrons in scattering states as non-interacting, allows us to integrate them out exactly.
Their influence on the $d$-level can be encapsulated in a self-energy function~\cite{Hewson}, whose imaginary part
is  $\pi |V|^2 \tilde{\rho}_{B}(\epsilon)$,  where 
$\tilde{\rho}_B(\epsilon) = \rho_B(\epsilon \neq 0)$,  
being $\rho_B(\epsilon)$   the local density of states at the apical atom~\cite{suppl}.
The latter exhibits a singular behavior $ \rho_B(\epsilon) \sim \frac{1}{|\epsilon| \ln^2|\epsilon/D|}$  ($D\sim t$) 
around the DP~\cite{Pereira2006Nuno}.
We can mimic the effect of the self-energy by representing the $\pi$ band as a electron reservoir with 
energy-dependent tunneling into the $d$-level~\cite{Hewson}, leading to 
\begin{align}
&H = H_V +  H_{\mathrm{hyb}} +  \int^{+D}_{-D} d\epsilon \, \epsilon \: \psi^{\dag}_{\alpha}(\epsilon)\psi^{\alpha}(\epsilon),\notag\\
&H_{V} =  \epsilon_d n_d + \notag U n_{d\uparrow} n_{d\downarrow}  +   \epsilon_D n_0 
  + U_{\sigma \pi} n_d n_0 - J_H \mathbf{S}_d  \cdot \mathbf{S}_0,\\
&H_{\mathrm hyb} = V \int^{+D}_{-D} d\epsilon \, \left[ t(\epsilon) d^{\dag}_{\alpha} \psi^{\alpha}(\epsilon) + t^*(\epsilon)\psi^{\dag}_{\alpha}(\epsilon) d^{\alpha}\right], \label{eq:am}
\end{align}
where $\epsilon_D = 0$, $\Delta(\epsilon) =  |t(\epsilon)|^2 = \tilde{\rho}_B(\epsilon)$, and
$D=D_0\sim t$ is a high-energy cut-off of order of the width of the $\pi$ band. Since (for small angle $\theta$) 
$\Delta(\epsilon_d)/|\epsilon_d| \ll 1$,  charge fluctuations in the dangling $\sigma$ orbital are small 
at low temperatures. However, the charge state 
of the ZM is determined by the magnitude of the parameter $G =J_H/4 U_{\sigma\pi}$: $\langle n_d\rangle \approx 0$ (doublet)
for $G\ll 1$ ,  and $\langle n_d \rangle \approx 1$ (triplet) for   $G\gg 1$. In both cases,
 the charge fluctuations can be integrated out using a Schrieffer-Wolff transformation, which leads to a Kondo model:
\begin{equation}
H_K = H_c + J_K\: \mathbf{s}_0 \cdot \mathbf{S}_{V} + V_0 f^{\dag}_{\alpha 0} f^{\alpha}_0,\label{eq:km}
\end{equation}
where $\mathbf{s}_0 = f^{\dag}_{\alpha 0} \left(\frac{\boldsymbol{\sigma}}{2} \right)^{\alpha}_{\beta} f^{\beta}_0$, $f^{\alpha}_0 = 
\int  d\epsilon \, t(\epsilon)\: \psi^{\alpha}(\epsilon)$  and 
$\mathbf{S}_V$ is the \emph{total} spin operator of the vacancy: $\mathbf{S}_{V} = \mathbf{S}_d$, for the doublet
($G \ll 1$) and    $\mathbf{S}_{V} = \mathbf{S}_d + \mathbf{S}_0$, for the triplet ($G \gg 1$). The
expressions for  $J_K$ and the potential $V_0$  depend on the total spin of the vacancy~\cite{suppl}.

However,  for $G\approx 1$, the
charge fluctuations in the ZM level are not negligible and the description in terms of Eq.~\eqref{eq:km} breaks down.
In such a case, we must deal with the full Anderson model, Eq.~\eqref{eq:am}. Having $G\approx 1$
requires fine tuning as it corresponds to a quantum phase transition between the doublet and the triplet, which will be
analyzed elsewhere~\cite{unpub}. Here,  we assume that
$U_{\sigma\pi} |1-G| $ is large enough so that using  Eq.~\eqref{eq:km} with a 
cut-off $D_0 \lesssim \mathrm{min} \{ |\epsilon_d|, U_{\sigma \pi} |1-G|\}$ is sufficient.

Next, we carry out a scaling analysis  of~\eqref{eq:km}~\cite{Hewson,Fradkin,Anna}. 
The perturbative scaling equations for $j_K = J_K/D_0$
and $v = V_0/D_0$ read:
\begin{align}
\frac{dj_K}{d \ln D}  &= r(D) j_K  - j^2_K + O(j^3_K), \,
 \frac{d v}{d\ln D} &=  r(D) v, \label{eq:rg}
\end{align}
 where $r(D) = (d\ln \Delta(\epsilon)/d\ln \epsilon)_{\epsilon=D}$ is assumed to be a slowly
 varying function of $D$, which is the case since near the DP $\Delta(\epsilon) \sim [|\epsilon| \ln^2 (\epsilon/D)]^{-1}$ 
 and   $r(D) \simeq -1$ with logarithmic accuracy. Thus, as $D$ is reduced, $j_K(D)$ and $v(D)$ grow
 and the vacancy enters the strong coupling regime. For $j_K(D)$, 
 a similar  conclusion was reached in~\cite{Hentschel} based on an analysis of the 
 orthogonality  catastrophe in defective graphene.  In the strong coupling regimem where $j_K(D^*) \sim 1$ (or $v(D^*) \sim 1$), 
 (\ref{eq:rg}) break down. However, the Kondo temperature can be related to 
 the crossover  scale $D^*$ where $j_K(D^*) \sim 1$  (provided $j_K\sim 1$ before $v$).  The solution of~\eqref{eq:rg}
yields  $T_K \ln^2 (T_K/D_0) \sim J_K$  for $J_K\ll D_0$~\cite{suppl}. 
Note that, this means a dramatic enhancement of $T_K$ compared to the 
vanishing (exponentially small) $T_K$ of the 
pseudo-gap~\cite{Fradkin,SenguptaBaskaran,Uchoa2011,Anna}  (flat-band~\cite{Hewson}) 
 Kondo models.

  In the strong coupling regime, we have to rely on other methods different from Eqs.~(\ref{eq:rg}).  
In this regard,  there are remarkablesimilarities of the present Kondo model 
with one studied by Bulla and Vojta ~\cite{BullaVojta} using the numerical renormalization group (NRG).
These authors considered a spin-$1/2$ Kondo model where $\Delta(\epsilon) \sim |\epsilon|^r$, with  
$-1< r  < 0$.  With logarithmic accuracy, we can borrow their results for $r \downarrow -1$  and 
$J_K> 0$. Thus,  two stable strong coupling fixed-points (FPs) are known to exist: a 
particle-hole asymmetric local moment  (ALM) FP, at which the vacancy magnetic moment is decouples from the band, and a 
singlet strong-coupling fixed point (SSC) FP, at which the single electron in dangling $\sigma$ orbital forms a
singlet with the conduction band electrons, and vacancy magnetic moment is thus quenched.  The flow between ALM and 
SSC FPs is controlled by a critical point~\cite{BullaVojta}.
\begin{table}[t]
\begin{center}
\begin{tabular}{c|c|c}
\hline
Vacancy Spin & $j_k\sim 1$ & $v\sim 1$\\
\hline
Doublet ($S_V = \frac{1}{2}$) & SSC  & ALM \\
Triplet ($S_V = 1$) & USSC & ALM\\ 
\hline
\end{tabular}
\end{center}
\caption{Stable phases for the Kondo model of Eq.~\eqref{eq:km}. SSC stands for singlet strong coupling ,
 ALM for asymmetric local moment, and USSC for underscreened strong coupling; 
 $j_K\sim 1$ and $v\sim 1$ indicate  which coupling (either $j_K \propto J_K$ or $v\propto V_0$) in
Eq.\eqref{eq:km} becomes relevant first.}
\label{table1}
\end{table}

  For the triplet case ($G \gg 1$), to the best of our knowledge, there are no NRG results available. However, in this case 
we can tentatively assume that,  as $j_K$ flows to strong coupling, half of the vacancy spin-$1$ will be quenched. 
Applying an argument due to  Nozi\'eres and Blandin~\cite{NozieresBlandin}, the residual  spin-$1/2$ couples
ferromagnetically (FM)  to the conduction band, that is, for $T \ll T_K$, it is
described by $H_{\mathrm{SC}} = H_c  + J^{\prime} \mathbf{s}_1 \cdot \mathbf{S}^{\prime}_{V}$
where $\mathbf{S}^{\prime}_V$ is the residual vacancy spin, 
and $J^{\prime} < 0$. In this strong-coupling coupling picture, one electron is captured 
at the first site (site $0$ in our notation above) of the 
Wilson chain~\cite{Hewson}  representation of $H_c$ (cf. Eq.~\ref{eq:am}), thus projecting out this site.
The residual spin-$1/2$  couples  to the next site of the Wilson chain, whose 
spin is described by the operator $\mathbf{s}_1$. The hybridization function for this site
can be obtained by tridiagonalization, which allows to relate the local Green's functions at sites $0$ 
and $1$ with  $0$ site projected out  (see~\cite{Hewson}, chap. 4):
$g_1(\epsilon) = \frac{\epsilon}{|\tilde{V}|^2} - \frac{1}{|\tilde{V}|^2 g_0(\epsilon)}$, 
with $\tilde{V}\propto V$. Hence
$\Delta_1(\epsilon) \propto - \mathrm{Im}\: g_1(\epsilon) \sim |\epsilon|/\ln^2|\epsilon/D|$. Hence, with logarithmic
accuracy,  $r(D) \simeq 1$ in Eq.~\eqref{eq:rg}, which, together with $j_K(0) \sim J^{\prime} < 0$,
implies that  $J^{\prime}$ renormalizes to zero. Thus, 
 $H_{SC}$ describes a stable \emph{underscreened Kondo} FP. Moreover, as in the double case, 
 for large particle-hole asymmetry, $V_0$, it is also reasonable to expect the existence of a strong-coupling fixed similar 
 to the  ALM FP, and a critical point in the $V$-$J_K$ plane that controls the flow between the two  FPs. 
 The possible phases of \eqref{eq:km} are summarized in Table~\ref{table1}.

 Finally, let us consider the effect and the next
neighbor hopping, $t^{\prime}$, and a small doping $\mu\neq 0$. At first sight,   both perturbations introduce 
particle-hole asymmetry~\cite{Pereira2006Nuno}, which, for $\mu, t^{\prime} \ll t$, 
can be regarded as additional contributions to the bare value of $V_0$ in Eq.~\eqref{eq:km}.
Thus, the initial conditions for the scaling equations~\eqref{eq:rg} are modified from their 
Schrieffer-Wolff values~\cite{suppl}, which favors the flow towards the ALM FP.  
For $\mu\neq 0$, the flow towards the 
ALM FP can be seen as the first stage of a two-stage flow~\cite{Vojta2010},  whose final stage occurs at much 
lower temperatures and  may be a crossover from the ALM  to a more  conventional Kondo-singlet FP (see below). 
The RG flow for the triplet case will be analyzed elsewhere~\cite{unpub}.

To test the above ideas and obtain some quantitative estimates for $T_K$, 
we have used a slave-boson mean-field approach to  Anderson model
in Eq.~\eqref{eq:am} in the $U\to +\infty$ limit~\cite{Hewson,NewnsRead}. In 
the mean-field theory~\cite{Hewson,NewnsRead}, $T_K$ is determined 
from the equation:
\begin{equation}
 T_K \sum_{\omega_n } \frac{g(i\omega_n)}{i \omega_n}e^{i\omega_{n}0^{+}} = \frac{1}{J_K},
\end{equation}
where $\omega_n = 2\pi (n + \frac{1}{2}) T_K$, 
$J_K = 2 |V|^2/|\epsilon_d|$ is the 
Kondo coupling for $U\to+\infty$~\cite{suppl}, and $g(i\omega_n) = 
\int d\epsilon \Delta(\epsilon)/(i\omega_n - \epsilon +\mu)$
is the local Green's function at the apical atom. The results of a
numerical solution of this equation are displayed in Fig.~\ref{fig:fig2}.
From Fig.~\ref{fig:fig2}(a), it can be seen that turning on a finite $\mu$  or
$t^{\prime}$ strongly suppresses $T_K$. For finite $t^{\prime}$,  $T_K$ appears to vanish 
at small $J_K \ll T_K(J_K,\mu=0,t^{\prime}\neq 0)$, which
we interpret as  the ALM FP being favored over the SSC FP. 
However, the effect of  $\mu\neq 0$ seems to be slightly different because the curve for $T_K$ 
changes its concavity, suggesting a finite but small $T_K$ at small $J_K$ (finite-size effects
prevent us from obtaining accurate results for $T_K$ below $\sim 1$ K). Fig~\ref{fig:fig2}(b) displays
the values of $T_K$ for $t^{\prime}/t = 0.02$ (i.e. $t^{\prime} = 59$ meV) and several
values of $\mu$. We find that when $t$ has opposite sign to $\mu$,  their  
particle-hole symmetry-breaking contributions  appear to partially cancel 
each other, leading to a less pronounced suppression of $T_K$ (relative to the results in
Fig.~\ref{fig:fig2}(a)). We believe this is because a reduction of the bare value of $V_0$ in 
Eq.~\eqref{eq:km}  favors the SSC over the ALM FP for large enough  $J_K$.


In conclusion, we have studied a model for a reconstructed single-atom vacancy in graphene,
finding that the strong vacancy scattering potential can dramatically affect its magnetic
properties. We find that the, if  particle-hole symmetry breaking is not too strong,
Kondo effect can occur for small doping levels provided the dangling $\sigma$ orbital
is not passivated. However, for sufficiently large particle-hole
symmetry breaking, a local moment that becomes increasingly decoupled
from the  $\pi$ band as  temperatures decreases appears. Further consequences of 
our theory for existing and future experiments will be presented elsewhere~\cite{unpub}.

 We acknowledge illuminating discussions with V. Pereira, I. Grigorieva, D. Sanchez-Portal, and 
 E.~J. ~G. Santos. MAC and FG acknowlege the hospitality of KITP, Santa Barbara. This research has been supported by
ERC, grant 290846 and  MICINN, Spain, (FIS2008-00124, FIS2011-23713, CONSOLIDER CSD2007-
00010) (FG) and  FIS2010-19609-C02-02 (MAC), 
and partially supported by the National Science Foundation under Grant No. NSF PHY11-25915.
AI acknowledges financial support from CONICET (PIP 0662), 
ANPCyT (PICT 2010- 1907) and UNLP (PID X497), Argentina.

\newpage
\widetext
\begin{center}
\l arge \bf Supplementary Material
\end{center}

\setcounter{equation}{0}

\section{Details of the Model of the Vacancy}

\begin{figure}[b]
\includegraphics[scale=0.4]{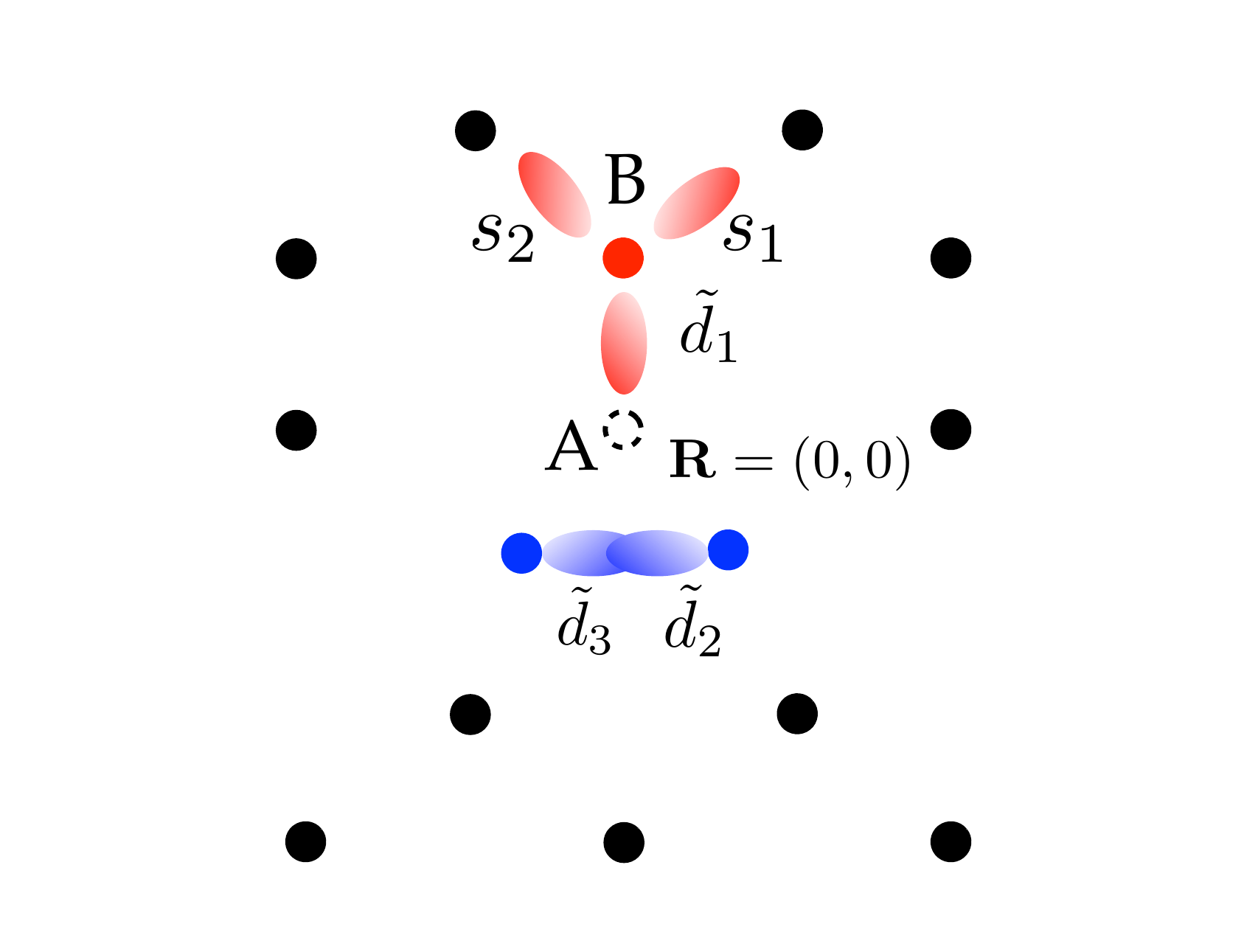}
\caption{Labeling of the relevant orbitals around the vacancy. We make the approximation that the vacancy $\sigma$-orbitals 
correspond essentially to the dangling bonds around the vacancy site (empty circle). See Eq.~\eqref{eq:ham1} for details.}
\label{fig:fig4} 
\end{figure}

\begin{figure}[ht]
\includegraphics[scale=0.3]{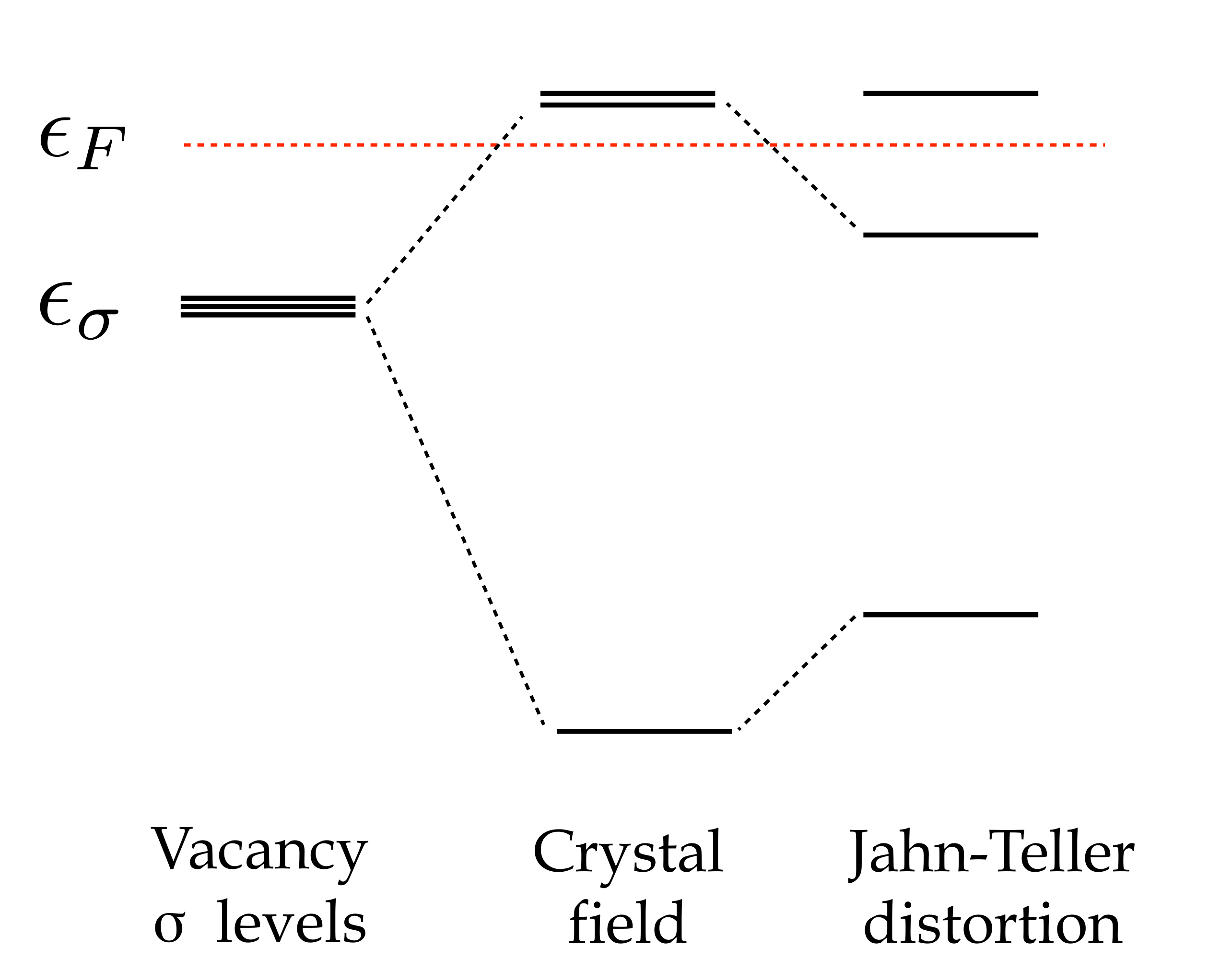}
\caption{Vacancy $\sigma$-levels and evolution when different perturbations are taken into account.  For left
to right: i) Crystal field splitting (i.e. nearest neighbor hopping $T=T^{\prime}$ and ii) Jahn-Teller distortion with $T > T^{\prime}$.
The lowest level, which has equal weight in the three atoms surrounding the vacancy has lowest energy and it is assumed
to be doubly occupied, thus forming a (stretched) $\sigma$ bound at the basis of a pentagon (cf. figure~\ref{fig:fig1} in the 
main text). The intermediate level is below the Fermi energy $\epsilon_F$ (assumed to be at the Dirac point) and we shall assume
it to be singly occupied as $U \approx 10$ eV~\cite{hubbardUb}}.
\label{fig:fig3} 
\end{figure}

  We  next describe the derivation of the model that is analyzed in the main text.  In establishing a minimal model for a
 reconstructed single-atom vacancy in graphene, we shall rely on existing first-principle calculations (e.g. Refs.~\onlinecite{Zunger1978b,Nanda2011b,dosSantos2011b,PalaciosYndurainb}). The results of those calculations can be summarized as follows:
\begin{enumerate}
\item
Removing one carbon atom (i.e. creating a vacancy) leaves three dangling $\sigma$-bonds behind.
The latter manifest themselves as three impurity levels splitting off the sigma band.  
\item
From a tight-binding perspective, the  vacancy $\sigma$-levels can be regarded as essentially the 
dangling $\sigma$ bonds of the atoms around the vacancy plus some admixture of $\sigma$
states away from the vacancy~\cite{Nanda2011b}.
\item
In an unrelaxed lattice, where all carbon atoms around the vacancy site remain at the same distance as 
the perfect graphene lattice, the crystal field effects, which correspond to  the 
nearest neighbor hopping between $\sigma$-orbitals in the original lattice) split the 
tree dangling-bond levels into one lower energy level (the symmetric combination)
and two degenerate excited levels (cf. Fig.~\ref{fig:fig3}). 
\item \label{jahnteller}
The  situation described in the previous point is energetically unstable 
against a Jahn-Teller distortion, which breaks the $D_{3h}$ symmetry 
of the unrelaxed lattice and draws two of the three carbon atoms surrounding the vacancy 
closer to each other (cf. Fig.~\ref{fig:fig1} in the main text and and Fig.~\ref{fig:fig4}).
\item
Further \emph{ab initio} calculations~\cite{dosSantos2011b} suggest that, in the presence of isotropic
compression, the three atoms around the impurity are no longer coplanar.  This is possible if e.g. the apical atom
(which contains a dangling $\sigma$--bond~\cite{Nanda2011b}) is lifted slightly above (or below) the plane of the two atoms 
opposite to it (i.e. those that form a reconstructed $\sigma$ bond as a result of the Jahn-Teller distortion), see Fig.~\ref{fig:fig1}
in the main text. This non-coplanarity is made it possible by  sp$^2$-sp$^3$ hybridization of the 
atoms around the vacancy. In the model below, we  assume that such hybridization is particularly 
important in the case of the apical atom (cf. Fig.~\ref{fig:fig1} in the main text). This allows the electrons to hop in and out of
the  dangling $\sigma$ orbital at the apical site. This  effect can be also achieved by rippling that occurs at any
finite temperature.
\end{enumerate}

 Given the above considerations, we propose the following model of the vacancy:~\footnote{We assume that the atom
that has been removed belongs to the $A$ sublattice. In other words, that $B$ is the majority sublattice.}
\begin{align}
H &= H_{\pi + V} + H_{\sigma} + H_{V} + H_{\mathrm{hyb}} + H_{d\pi},  \label{eq:ham1} \\
H_{\pi + V} &= - t \sum_{\langle i, j \rangle, \alpha} \left[ a^{\dag}_{\alpha i} b^{\alpha}_j  + b^{\dag}_{\alpha j} a^{\alpha}_{j} \right] 
+ \epsilon_0  \sum_{\alpha} a^{\dag}_{\alpha 0} a_{\alpha 0}, \label{eq:hampi}\\
H_{V} &=   \epsilon_{\sigma} \sum_{\alpha,i=1}^3 \tilde{d}^{\dag}_{\alpha i} \tilde{d}^{\alpha}_i - T \sum_{\alpha}
 \left[ \tilde{d}^{\dag}_{\alpha 2} \tilde{d}^{\alpha}_3 + \tilde{d}^{\dag}_{\alpha 3} \tilde{d}^{\alpha}_2 \right] 
  - T^{\prime} \sum_{\alpha} \left[  \tilde{d}^{\dag}_{\alpha 1} \tilde{d}^{\alpha}_2 +  \tilde{d}^{\dag}_{\alpha 2} \tilde{d}^{\alpha}_{1}
  +  \tilde{d}^{\dag}_{\alpha 1} \tilde{d}^{\alpha}_3 +  \tilde{d}^{\dag}_{\alpha 3} \tilde{d}^{\alpha}_{1} \right]
 + H_{V-\mathrm{int}}, \\
H_{\mathrm{hyb}} &=   \tilde{V}_{\pi\sigma} \sum_{\alpha} \left[ b^{\dag}_{\alpha 0} \tilde{d}^{\alpha}_1 +
 b^{\dag}_{\alpha 0} \tilde{d}^{\alpha}_1  \right]  + \tilde{V}_{\pi \sigma} \sum_{\alpha}  b^{\dag}_{\alpha 0}
 \left[ b^{\dag}_{\alpha 0} s^{\alpha}_{1} +  b^{\dag}_{\alpha 0}  s^{\alpha}_{2} +  
 s^{\dag}_{\alpha 1} b^{\alpha}_{0} +  s^{\dag}_{\alpha 2}  b^{\alpha}_{2}   \right],\label{eq:hyb}
\end{align}
where $T > T^{\prime}$ is the hopping between the vacancy $\sigma$-orbitals and 
$\tilde{V}_{\pi \sigma}$ describes the sp$^2$-sp$^3$ hybridization~\cite{CastroNetoGuineab} 
of the apical atom (see Fig.~\ref{fig:fig4} for details about the labeling of the different orbitals):
\begin{equation}
\tilde{V}_{\sigma\pi} = A \sqrt{\frac{1-A^2}{3}} (\epsilon_{s}-\epsilon_{p}),
\end{equation}
where $\sin \theta = A/\sqrt{A^2+2}$,  $\theta$ being the angle between the $\sigma$ orbitals at the apical
atom site (cf. Fig.~\ref{fig:fig4}) and the Graphene plane, and $\epsilon_s-\epsilon_p = -8.31 eV$ is the atomic 
s-p splitting. The above model neglects the local 
modifications of the hopping parameters around the vacancy site, both resulting from the Jahn-Teller
distortion and from the sp$^2$-sp$^3$ hybridization caused by an isotropic compression. 
In Eq.~\eqref{eq:hyb} $H_{d\pi}$ describes the Coulomb interaction between the 
$\sigma$ and the $\pi$ orbitals, which is particularly strong for the $\pi$ and $\sigma$  orbitals localized
around the vacancy (see Sec.~\ref{sec:twolevel}).
 
Furthermore, In Eq.~\eqref{eq:ham1}, the term $H_{V-\mathrm{int}}$ describes the interactions 
between the electrons localized at the  vacancy-$\sigma$ orbitals, which we will specify below. 
In the above equation, the limit  $\epsilon_0\to +\infty$ is implicitly assumed in order to remove 
the site  at $\mathbf{R} = (0,0)$ of the $A$ sublattice. 

 As discussed in point~\ref{jahnteller} above, two of the dangling bonds bind when the lattice
 undergoes a Jahn-Teller distortion. If we diagonalize the subspace spanned by the vacancy $\sigma$ 
 orbitals by means of the transformation:
\begin{align}
\left(\begin{array}{c}
d^{\alpha}_1\\
d^{\alpha}_{2}\\
d^{\alpha}_{3}
\end{array}
 \right) = 
 \left( \begin{array}{ccc}
u_{11} & u_{12} & u_{13}\\
u_{21} & u_{22} & u_{23}\\
u_{31} & u_{32} & u_{33}
 \end{array}
 \right) 
 \left(\begin{array}{c}
\tilde{d}^{\alpha}_{1}\\
\tilde{d}^{\alpha}_{2}\\
\tilde{d}^{\alpha}_{3}
\end{array}
 \right) 
\end{align}
where $\mathbf{u}_{1} = (u_{11},u_{12},u_{13}) =  \mathcal{N}_1 (-\frac{1 + \sqrt{1+ 8 r^2}}{2 r},1,1)$,
$\mathbf{u}_{2} = (u_{21},u_{22},u_{23}) =  \mathcal{N}_2 (\frac{1 - \sqrt{1+ 8 r^2}}{2 r},1,1)$,
$\mathbf{u}_{3} = (u_{31},u_{32},u_{33}) = \frac{1}{\sqrt{2}}  (0,1,-1)$,  with $r = T^{\prime}/T$ and
$\mathcal{N}_{i}$ the normalization constants. The energies of the levels resulting from the 
previous diagonalization are given by first term in
\begin{equation}
H_{V} = \sum_{\alpha, i=1}^3 \epsilon_{i} d^{\dag}_{\alpha i} d^{\alpha}_{i} + H_{\mathrm{V-int}},
\end{equation} 
where $\epsilon_1 =\epsilon_{\sigma} +  \frac{1}{2}T \left(\sqrt{1+8r^2}-1\right)$, $\epsilon_2 =\epsilon_{\sigma} -  \frac{1}{2}T \left(\sqrt{1+8r^2}+1\right)$, $\epsilon_3 = \epsilon_{\sigma} + T$ (According to the fitting to DFT-LDA calculations carried out by 
Nanda \emph{et al.} in Ref.~\onlinecite{Nanda2011b}, $T = 1.85$ eV and $T^{\prime} = 1.30$ eV). 
Thus, the level number $3$ corresponds
to an excited state, which lies above the Dirac point,~\cite{Nanda2011b} and  the level $2$ corresponds to a deep
level describing the  reconstructed bond at the basis of the pentagon opposite to the apical atom (cf. 
Fig.~\ref{fig:fig1} in the main text and \ref{fig:fig4}).   

 The remaining electron from the vacancy $\sigma$ orbitals  occupies 
the orbital with energy $\epsilon_1$, and in flat graphene it cannot not hybridize with the $\pi$-band although it has a
very large amplitude at the apical atom (see expression for $\mathbf{u}_1$ above).
However, in rippled graphene (as in the presence of isotropic compression) there is hybridization, 
 through the term $H_{\mathrm{hyb}}$ in Eq.~\eqref{eq:ham1}.
In the following, we shall investigate whether such hybridization could lead to a Kondo effect.  Thus, we shall focus 
on the orbital $1$ and, to lighten the notation, henceforth we shall drop the orbital index (the other two orbitals, $2$ and $3$,
will be treated as inert). Furthermore, we shall neglect the last  term in Eq.~\eqref{eq:hyb}, that is, the sp$^2$-sp$^3$
hybridization with the $s_1$ and $s_2$ $\sigma$-orbitals of the apical atom (cf. Fig.~\ref{fig:fig4}). Such terms are
expected to lead to a renormalization of the hybridization of the vacancy level $1$ with the $\pi$ band. These
considerations lead to the following  Anderson model:
\begin{equation}
H_{\mathrm{AM}} = \epsilon_d \sum_{\alpha} d^{\dag}_{\alpha}d_{\alpha} + V\sum_{\alpha} \left[ b^{\dag}_{\alpha 0} d^{\alpha} +
d^{\dag}_{\alpha} b^{\alpha}_{0} \right] + U d^{\dag}_{\uparrow} d^{\dag}_{\downarrow} d^{\downarrow} d^{\uparrow} + H_{\pi + V}\label{eq:hamam} + H_{d\pi}
\end{equation}
where $H_{\pi+V}$ is given by Eq.~\eqref{eq:hampi} with $\epsilon_0\to +\infty$; 
$\epsilon_d = \epsilon_1 = \epsilon_{\sigma} + \frac{1}{2}T \left(\sqrt{1+8r^2}-1\right)\simeq -0.75$ eV,~\cite{Nanda2011b,PalaciosYndurainb} and $V = |\varphi_{d}(\boldsymbol{0})|^2 V_{\sigma \pi}(\theta)$, where $|\varphi_{d}(\boldsymbol{0})|^2 \approx 0.7$, according to the above tight-binding model, and $U \approx 10$ eV~\cite{hubbardUb}.
 
The novelty of \eqref{eq:hamam} lies in the fact that it is a 
single-impurity Anderson model in the presence of a strong scattering potential that induces a resonance
at the Dirac point. Note that in the above model we have neglected any interactions between
the electrons in the $\pi$-band.

\begin{figure}[hb]
\includegraphics[scale=0.4]{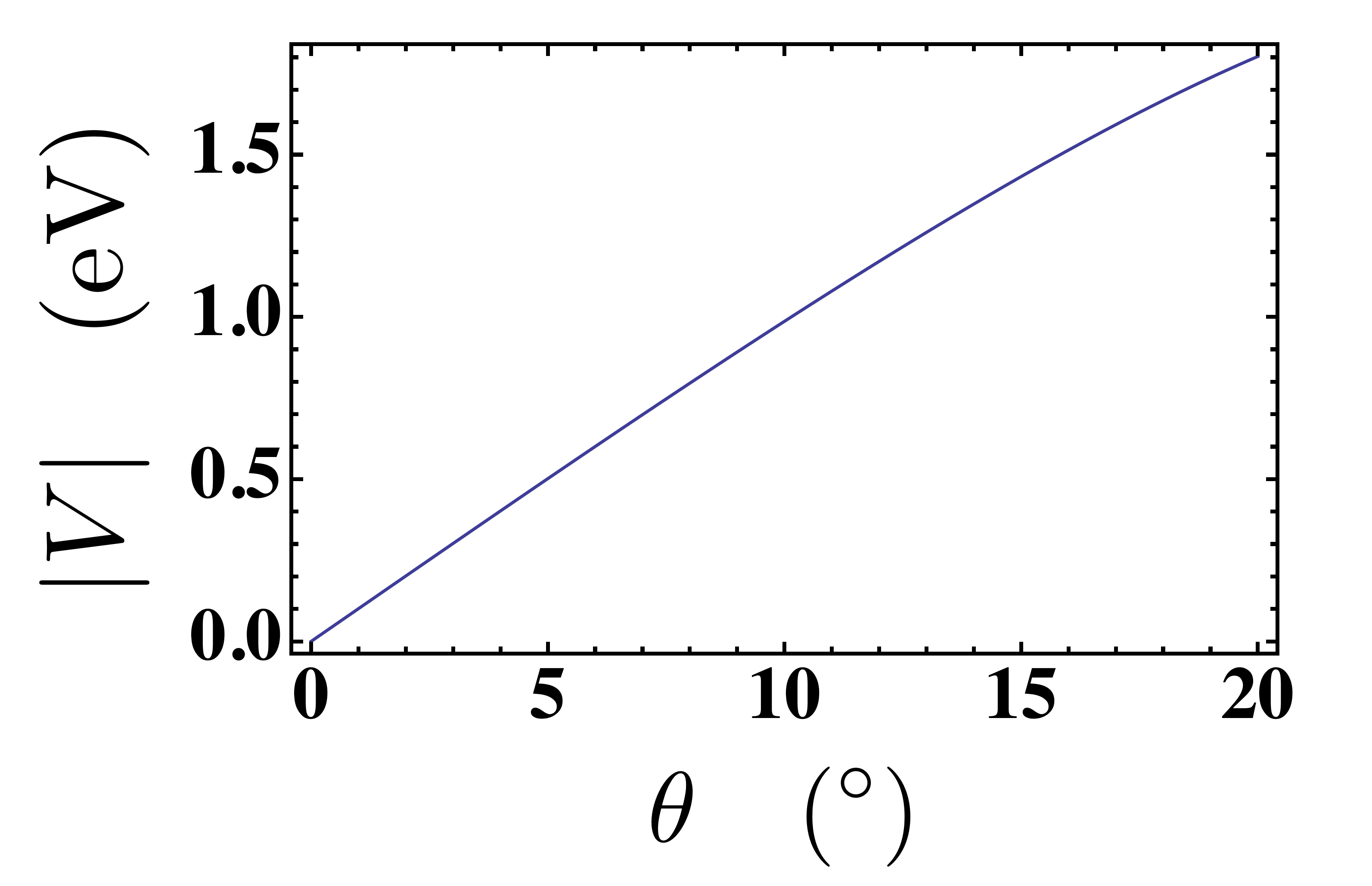}
\caption{sp$^2$-sp$^3$ hybridization of the dangling $\sigma$  orbital in the apical atom (cf. Fig.~\ref{fig:fig1} in the main text) 
with the $\pi$ orbitals on the same atom vs. angle between the dangling orbital and the Graphene plane.}
\label{fig:fig5}
\end{figure}

\section{Two level Anderson model}\label{sec:twolevel}

 Let us return to the above Hamiltonian and consider first the scattering problem of the $\pi$ band 
 electrons with the vacancy. Let us assume that we have  diagonalized the Hamiltonian \eqref{eq:hampi}
 and therefore we can write:
 \begin{equation}
 H_{\pi + V} = \varepsilon_0 \pi^{\dag}_{\alpha 0} \pi^{\alpha}_{0} + \sum_{\alpha,p,\mathbf{n}} \epsilon_{p \mathbf{n}} \:  
 \pi^{\dag}_{p \alpha \mathbf{n}} \pi^{\alpha}_{p\mathbf{n}},
 \end{equation}
where the index $p = c, v$ describes the valence and conduction band scattering channels and $\alpha = \uparrow,\downarrow$. 
The other quantum numbers have been collectively denoted by $\mathbf{n}$. We have explicitly separated the zero mode, 
whose energy coincides with the Dirac point energy ($\varepsilon_0 = 0 $ for a particle-hole symmetric case where we neglect
next nearest neighbor hopping),  and  it described by $\pi^{\dag}_{\alpha}, \pi^{\alpha}$. The original lattice operators can be 
expressed in terms of the eigenmodes of $H_{\pi+V}$ as follows:
\begin{align}
a^{\alpha}_{i} &=   \sum_{p,\mathbf{n}} \varphi_{A p\mathbf{n}}(\mathbf{R}_i) \: \pi^{\alpha}_{p\mathbf{n}},\\
b^{\alpha}_{i} &= \varphi_{0}(\mathbf{R}_i) \pi^{\alpha}_0 + \sum_{p, \mathbf{n}} \varphi_{B p\mathbf{n}}(\mathbf{R}_i)\: \pi^{\alpha}_{p\mathbf{n}}.
 \end{align}
In the last expression above, we have used the fact that the zero-mode wavefunction is (within the tight-biding nearest-neighbor
hopping approximation) is fully contained within the majority $B$ sublattice (see e.g. Ref.~\onlinecite{Nanda2011} and references therein).

 In this eigenmode basis, the minimal Anderson Hamiltonian, Eq.~\eqref{eq:hamam}, introduced in previous section reads:
\begin{align}
H_{\mathrm{AM}} &= H_{\pi + V} +  \epsilon_d  \sum_{\alpha} d^{\dag}_{\alpha} d^{\alpha} + U d^{\dag}_{\uparrow} d^{\dag}_{\downarrow}
d^{\downarrow} d^{\uparrow} + V \sum_{\alpha} \left[ \varphi_{0}(\boldsymbol{0}) d^{\dag}_{\alpha} \pi^{\alpha}_{0} + \varphi^{*}_{0}(\boldsymbol{0})    \pi^{\dag}_{\alpha 0} d^{\alpha} \right]\nonumber\\
&\quad\quad + V \sum_{p,\mathbf{n}} \left[  \varphi_{Bp\mathbf{n}}(\boldsymbol{0})  \: d^{\dag}_{\alpha} \pi^{\alpha}_{p\mathbf{n}} +
\varphi^*_{B p \mathbf{n}}(\boldsymbol{0}) \pi^{\dag}_{\alpha p \mathbf{n}}  d^{\alpha} \right] + H_{d\pi}\\ 
\end{align}
Note that the zero-mode wavefunction is not normalizable since (neglecting oscillatory terms) 
$|\varphi_{0}(\mathbf{R})|^2 \sim  \frac{N_0^2}{|\mathbf{R}|^2}$, which 
implies that
\begin{equation}
 \sum_{\mathbf{R}}  |\varphi_{0}(\mathbf{R})|^2  \simeq  N_0^2 \int d\mathbf{R}\,  \frac{1}{|\mathbf{R}|^2} \sim 2\pi N_0^2 \, \ln L,
\end{equation}
diverges as the linear size of the system $L \to +\infty$. If we insist in normalizing the zero-mode wavefunction  to unity, the normalization  constant $N_0\simeq (2\pi \ln L)^{-1/2} \to 0$ as $L\to +\infty$. This implies that the hybridization of the zero-mode with the localized
$\sigma$ level  scales as $\sim  (\ln L)^{-1/2}$. Though this vanishes in the thermodynamic limit, it does so very slowly.
For instance, for a typical experimental device $a_0\times L~ 1000$ nm. Since $a_0 \simeq 0.25$ nm, $L \simeq 4000$,which
implies that $N_0^2\simeq 1/50 = 0.02$ (i.e. $N_0\simeq 0.1$).

 At this point, it is worth considering also the interaction terms between the electrons in the dangling $\sigma$ level  and the 
electrons in the $\pi$ band. Those terms result from the Coulomb interaction,
\begin{equation} 
H_C = \frac{e^2}{2\epsilon}\int d\mathbf{r} d\mathbf{r}^{\prime} \,  \frac{\rho(\mathbf{r}) \rho(\mathbf{r}^{\prime})}{|\mathbf{r}-\mathbf{r}^{\prime}|},\label{eq:coul}
\end{equation}
where $\rho(\mathbf{r}) = \Psi^{\dag}_{\alpha}(\mathbf{r}) \Psi^{\alpha}(\mathbf{r})$. If we restrict ourselves to the subspace
spanned by the $\sigma$ level and the $\pi$  and zero mode, then we can approximate:
\begin{equation}
\Psi^{\alpha}(\mathbf{r}) \approx \varphi_{d}(\mathbf{r}) d^{\alpha} + \varphi_{0}(\mathbf{r}) \pi^{\alpha}_0.
\end{equation}
Hence, 
\begin{equation}
\rho(\mathbf{r}) =  |\varphi_d(\mathbf{r})|^2 d^{\dag}_{\alpha} d^{\alpha} +  |\varphi_0(\mathbf{r})|^2 \pi^{\dag}_{\alpha} \pi^{\alpha}_0
+ \varphi^{*}_d(\mathbf{r}) \varphi_0(\mathbf{r}) d^{\dag}_{\alpha} \pi^{\alpha}_0
+ \varphi^{*}_0(\mathbf{r}) \varphi_d(\mathbf{r}) \pi^{\dag}_{\alpha} d^{\alpha}_0
\end{equation}
Upon substitution of the above  in Eq.~\eqref{eq:coul}, and upon neglecting terms that involve inter-level transitions, we obtain:
\begin{equation}
H_{\mathrm{int}} =  \tilde{U}_{\sigma\pi} n_d n_0  + U_{\pi\pi} n^2_0 - J_H d^{\dag}_{\alpha} \pi^{\dag}_{\beta 0} \pi^{\alpha}_{0} d^{\beta},
\end{equation}
where we have subtracted the self-interaction (i.e. Hubbard-$U$) term for the $\sigma$ level ($\sim U n^2_d$).
\begin{align}
\tilde{U}_{\sigma\pi} &= \frac{e^2}{\epsilon} \int d\mathbf{r}\,  |\varphi_{d}(\mathbf{r})|^2 |\varphi_{0}(\mathbf{r^{\prime}})|^2 \frac{1}{|\mathbf{r}-\mathbf{r}^{\prime}|}, \label{eq:usp}\\ 
U_{\pi\pi} &= \frac{e^2}{\epsilon} \int d\mathbf{r}\,  |\varphi_{0}(\mathbf{r})|^2 |\varphi_{0}(\mathbf{r^{\prime}})|^2 \frac{1}{|\mathbf{r}-\mathbf{r}^{\prime}|},\label{eq:upp}\\
J_H &=  \frac{e^2}{\epsilon} \int d\mathbf{r}  \, \varphi^*_{d}(\mathbf{r}) \varphi^*_{0}(\mathbf{r})  \varphi_{0}(\mathbf{r^{\prime}}) \varphi_{d}(\mathbf{r}^{\prime}) \frac{1}{|\mathbf{r}-\mathbf{r}^{\prime}|}.\label{eq:jh}
\end{align}
We also recall that, on dimensional grounds,  
\begin{equation}
U \sim \frac{e^2}{\epsilon a_0}
\end{equation}
Before trying to estimate the above Coulomb integrals, we note that, by using the following identity for the Pauli matrices
\begin{equation}
\sum_{a=x,y,z} \, \left( \sigma^a\right)^{\alpha}_{\beta} \left(\sigma^a \right)^{\mu}_{\nu} = 2\delta^{\alpha}_{\nu} \delta^{\mu}_{\beta} - 
\delta^{\alpha}_{\beta} \delta^{\mu}_{\nu},\label{eq:sigmaiden}
\end{equation}
we can recast
\begin{equation}
d^{\dag}_{\alpha} \pi^{\dag}_{\beta 0} \pi^{\alpha}_{0} d^{\beta} =  2\mathbf{S}_{d}\cdot \mathbf{S}_0  + \frac{1}{2} n_d n_0,
\end{equation}
where $S^a_0 = \pi^{\dag}_{0\alpha} \left( \frac{\sigma^a}{2}\right)^{\alpha}_{\beta} \pi^{\beta}_0$ and $S^a_V = d^{\dag}_{0\alpha} \left( \frac{\sigma^a}{2}\right)^{\alpha}_{\beta} d^{\beta}_0$. Thus, 
\begin{equation}
H_{d\pi} =   U_{d\pi} n_d n_0  + U_{\pi\pi} n^2_0 - J_H  \mathbf{S}_{d} \cdot \mathbf{S}_0,
\end{equation}
where 
\begin{equation}
U_{\sigma\pi} = \tilde{U}_{d\pi} - \frac{J_H}{2}
\end{equation}
Let us next find the scaling of the different couplings  $U_{\pi\pi}, \tilde{U}_{\sigma d}$, and $J_{H}$ with the system size.
To this end, we recall that (see e.g. Ref.~\onlinecite{Nanda2011}):
\begin{align}
\varphi_0(\mathbf{r}) &= N_0 \frac{1}{|\mathbf{r}|} \cos ((\mathbf{K}+\mathbf{K}^{\prime})\cdot \mathbf{r})  \sin ((\mathbf{K}-\mathbf{K}^{\prime})\cdot \mathbf{r} + \theta_{\mathbf{r}}),\\
\varphi_d(\mathbf{r}) &= N_d \frac{e^{-\lambda_d |\mathbf{r}|}}{|\mathbf{r}|},
\end{align}
where $\lambda_d \simeq a^{-1}_0$. We shall not perform the integrals in Eqs.~(\ref{eq:usp},\ref{eq:upp},\ref{eq:jh}) explicitly but merely note that, from the dependence on $\varphi_0(\mathbf{r})$ (i.e. on $N_0$) and, on
dimensional grounds, 
\begin{align}
U_{\pi\pi} &\sim \frac{e^2}{\epsilon a_0} \left(\frac{1}{2\pi \ln L} \right)^2,\\
J_H &\sim  \frac{e^2}{\epsilon a_0}  \frac{1}{2\pi \ln L},\\
U_{\sigma\pi}\sim \tilde{U}_{d\pi} &\sim  \frac{e^2}{\epsilon a_0}  \frac{1}{2\pi \ln L},
\end{align}
Note that $U_{\pi\pi}$ scales faster to zero as $L\to +\infty$ than the other couplings because the zero-mode self-interaction
is strongly suppressed by the fact that its wave-function is only marginally localized. 
Taking $\frac{e^2}{\epsilon a_0} \simeq10 $ eV, and 
$L^1\simeq 10^7$, we obtain: 
\begin{align}
U_{\pi\pi} &\sim 1\, \mathrm{meV},\\
U_{\sigma \pi} &\sim 100 \, \mathrm{meV},\\
J_{H} &\sim 100 \, \mathrm{meV},\\
U &\sim 10 \, \mathrm{eV}.
\end{align}
However, the hybridization energy of the dangling $\sigma$ orbital with the zero mode is given by $V N_0$. 
For an angle $\theta \simeq 2$ degrees, $|V|\simeq 200$ meV, and $N_0 = 0.1$ for $L \simeq 4000$, $|V| N_0 \simeq 20$ meV.
Indeed, whereas the Hund interaction $-J_{H} \mathbf{S}_d \cdot \mathbf{S}_{0}$ tends to align the spins of the $\sigma$ level
and the $\pi$ zero mode ferromagnetically, the hybridization $V N_0$ favors  an anti-ferromagnetic alignment, but since 
$J_H \gg V^2/|\epsilon_d + U_{\sigma\pi}| \simeq 1$ meV, we can safely assume that the triplet (i.e. the ferromagnetism) is favored 
provided the  each level is singly occupied, i.e. $n_d + n_0 = 2$. On the other hand, the interaction term $\propto U_{\pi\sigma}$ tends to favor  the emptiness of one of the levels, i.e. $n_d + n_0 = 1$, which implies that a spin doublet  is favored. In the following 
section, we analyze the different possibilities for the total spin of the vacancy. 

 The above estimates account for the contribution from the direct Coulomb interaction between the  levels localized at the 
vacancy. However, other contributions may also exist. For example, the hybridization with the conduction band of both levels 
resulting from a next-nearest neighbor hopping $t^{\prime}$  may also lead to an additional (RKKY-type) 
contribution to the Hund's rule  coupling $J_H$~\cite{Galpin}. Furthermore, a large enough $t^{\prime}$ also 
shifts energy and broadens the zero-mode, which lead to its disappearance into the continuum of scattering states. 
For the above reasons, in the main text we have explored the possible magnetic phases treating $J_H$ and $U_{\sigma \pi}$
as unknown parameters. 

\section{The `Atomic' limit of the vacancy} \label{sec:atomlimit}

 Let us momentarily set the hybridization with the rest of the $\pi$ band to zero and consider the spectrum of the 
coupled $\sigma$ and $\pi$ (zero-mode) levels. The latter are described by the following Hamiltonian:
\begin{equation}
H_{\pi\sigma} = \varepsilon_0 n_d + \epsilon_d n_d + U n_{d\uparrow} n_{d\downarrow} + U_{\sigma \pi} n_d n_0 
- J_{H} \mathbf{S}_d \cdot \mathbf{S}_{0} + \left[ V_{\sigma\pi} d^{\dag}_{\alpha} \pi^{\alpha}_0 + V^*_{\sigma\pi} 
\pi^{\dag}_{\alpha 0} d^{\alpha} \right],\label{eq:hamat}
\end{equation}
where we have introduced
\begin{equation}
V_{\sigma \pi} = V \varphi_0(\mathbf{0}) \simeq V N_0 \simeq  V \left( \frac{1}{2\pi \ln L} \right)^{1/2}.
\end{equation}

 In what follows, we are going to determine the ground state of the two level `atom' described by the Hamiltonian
 in Eq.~\eqref{eq:hamat}. To this end, we shall rely on the fact that both $n_{T} = n_0 + n_d$ and $\mathbf{S}^2_T = \left(
 \mathbf{S}_{0} + \mathbf{S}_d\right)^2 = S_T (S_T +1)$ and $S^z_T = S^z_d + S^z_o$ are good quantum numbers.

 \begin{enumerate}
 \item  {\bf Doublet:}  For $n_T = n_d + n_0 = 1$, $S_T = \frac{1}{2}$, $S^{z}_T = \pm \frac{1}{2}$, the Hilbert (sub)space of the system is spanned by:
 \begin{align}
 |1,S_z = \pm\frac{1}{2}\rangle &=  d^{\dag}_{\alpha} |0\rangle,\\
 |2,S_z = \pm\frac{1}{2}\rangle &=  \pi^{\dag}_{\alpha} |0 \rangle,\\
 \end{align}
 with $\alpha = \uparrow$  ($\alpha = \downarrow$) for $S^z_T = \frac{1}{2}$ ($S^z_T = -\frac{1}{2}$).
 The eigen energies are $ \epsilon_{\pm}= \frac{\epsilon_d + \varepsilon_0 \pm \sqrt{(\epsilon_d - \varepsilon_0)^2- 4 |V_{\sigma \pi}|^2}}{2}$.  Since $|\epsilon_d - \varepsilon_0| \gg 2 |V_{\sigma \pi}|$, we find that the lowest energy state in th sector has an energy
 \begin{equation}
 E_0(n_T = 1, S_T = \frac{1}{2}) \simeq \epsilon_d - \frac{|V_{\sigma \pi}|^2}{|\epsilon_d - \varepsilon_0|}
 \end{equation}
 \item
{\bf Singlet:}  For $n_T = n_d + n_0 = 2$, with $S_T  = S^z_T = 0$. In this case, the Hilbert space of the system is spanned by the following three states:
 \begin{align}
 |1\rangle &= \frac{1}{\sqrt{2}} \left[ d^{\dag}_{\uparrow} \pi^{\dag}_{0\downarrow}  - d^{\dag}_{\downarrow} \pi^{\dag}_{0\uparrow}\right]
 |0\rangle,\\
 |2 \rangle &= d^{\dag}_{\uparrow} d^{\dag}_{\downarrow} |0\rangle, \\
 |3\rangle  &= \pi^{\dag}_{0 \uparrow} \pi^{\dag}_{0\downarrow} |0\rangle.
 \end{align}
 In this subspace the Hamiltonian $H_V$ becomes a $3\times3$ matrix:
 \begin{align}
 H_{\sigma \pi}(n_T = 2, S_T= 0) = \left( 
 \begin{array}{ccc}
 \epsilon_{d} + \varepsilon_0 + U_{\sigma \pi} + \frac{3J_{H}}{4} & \sqrt{2} V_{\sigma \pi} & \sqrt{2} V^*_{\sigma\pi}\\
 \sqrt{2} V^*_{\sigma\pi} & 2\epsilon_d + U + \frac{3J_{H}}{4}  & 0 \\
 \sqrt{2} V_{\sigma \pi} & 0 & 2\varepsilon_0 + \frac{3 J_{H}}{4}
 \end{array}
 \right)
 \end{align}
 where we have used that $\mathbf{S}_d \cdot \mathbf{S}_0 = \frac{1}{2} \left[ \mathbf{S}^2_T - \mathbf{S}^2_d - \mathbf{S}^2_0 \right] = 
\frac{1}{2} \mathbf{S}^2_T - \frac{3}{4}$. The diagonalization of the above Hamiltonian is not possible explicity, but under the assumptions
that $\varepsilon_0 > \epsilon_d$ and $U\gg  |\epsilon_d|, |\varepsilon_d|$, the energy of the lowest state in this sector is: 
\begin{equation}
E_0(n_T = 2, S_T=0, S^z_T = 0) \simeq \epsilon_d + \varepsilon_0 - 2 |V_{\sigma \pi}|^2 \left[ \frac{1}{|\epsilon_d  + U - \varepsilon_0 - U_{\sigma \pi}|} + \frac{1}{|\epsilon_d + U_{\sigma\pi} - \varepsilon_0|} \right]
\end{equation}
\item
{\bf Triplet:} For $n_T = 2$, $S = 1$, this space is spanned by (note that the states in this case have different values of $S_z$, since
$H_V$ commutes with $S_z$, it  automatically becomes diagonal in this subspace):
\begin{align}
| S_T =  1, S^z_T = +1 \rangle &= d^{\dag}_{\uparrow} \pi^{\dag}_{0\uparrow} |0\rangle, \\
 | S_T = 1, S^z_T = 0\rangle &= \frac{1}{\sqrt{2}} \left[ d^{\dag}_{\uparrow} \pi^{\dag}_{0\downarrow}  + d^{\dag}_{\downarrow} \pi^{\dag}_{0\uparrow}\right]  |0\rangle,\\
 | S_T = 1, S^z_T = -1 \rangle  &= d^{\dag}_{\downarrow} \pi^{\dag}_{0\downarrow} |0\rangle.
\end{align}
 \begin{equation}
\langle S_T = 1, m |  H_{\sigma \pi} | S_T = 1, m^{\prime} \rangle =
\delta_{m,m^{\prime}}   \left( \epsilon_d + \varepsilon_0 + U_{\sigma\pi} - \frac{J_{H}}{4}\right)
 \end{equation}
Thus, the triplet consists of three degenerates states with energy:
\begin{equation}
E_0(n_T = 2, S_T = 1) = \epsilon_d + \varepsilon_0 + U_{\sigma\pi} - \frac{J_{H}}{4}.
\end{equation}
\item {\bf Triply occupied doublet}: 
For $n_T = 3$, $S = \frac{1}{2}$,  and $S^z = \pm \frac{1}{2}$ the Hilbert (subspace) is spanned by:
 \begin{align}
 |1,\alpha\rangle &=  d^{\dag}_{\alpha} \pi^{\dag}_{\uparrow 0}\pi^{\dag}_{\downarrow 0 } |0\rangle,\\
 |2,\alpha\rangle &=  d^{\dag}_{\uparrow} d^{\dag}_{\downarrow} \pi^{\dag}_{0 \alpha} |0 \rangle\\
 \end{align}
 To leading order, the energy of these states is given by:
 \begin{align}
 E_1 &= \langle 1, \alpha | H_{\sigma\pi} | 1, \alpha \rangle = 2\varepsilon_0 + \epsilon_d + 2 U_{\sigma \pi},\\
 E_2 &=\langle 2, \alpha | H_{\sigma\pi} | 2, \alpha \rangle = \varepsilon_0 + 2\epsilon_d + 2 U_{\sigma \pi} + U,
 \end{align}
 As in the double case, these energies are corrected by terms of order $|V_{\sigma \pi}|^2/|E_1-E_2|$ in the next leading order.
 
 \item {\bf Full vacancy singlet:} For $n_T = 4$, $S_T = S^z_T = 0$,  it is described by the state:
 \begin{equation}
 |\mathrm{Full}\rangle =  d^{\dag}_{\uparrow} d^{\dag}_{\downarrow} \pi^{\dag}_{\uparrow 0}\pi^{\dag}_{\downarrow 0 } |0\rangle,
 \end{equation}
 with energy $E_{\mathrm{full}} = 2\epsilon_d + 2\varepsilon_0 + 4 U_{\sigma \pi} + U$.
 \end{enumerate}

  Under the assumption that $U \simeq 10 \: \mathrm{eV} \gg \varepsilon_0, |\epsilon_d| \gg V_{\sigma\pi}$,  we must compare the spin doublet energy with the triplet energy in order to determine the spin state of the vacancy:
\begin{equation}
\Delta E_{\mathrm{triplet}-\mathrm{doublet}} = E_0(n_T = 2, S_T = 1) - E_0(n_T = 1, S_T = \frac{1}{2}) =   \varepsilon_0  + U_{\sigma \pi} - \frac{J_H}{4} + \frac{|V_{\sigma \pi}|^2}{|\epsilon_d  - \varepsilon_0|} 
\end{equation}
For example, if we choose the position of the chemical potential at the Dirac point and measure the energies
with respect to it,  $\mu = \varepsilon_0 = 0$, 
\begin{equation}
\Delta E_{\mathrm{triplet}-\mathrm{doublet}} = U_{\sigma\pi} - \frac{J_{H}}{4} + \frac{|V_{\sigma \pi}|^2}{|\epsilon_d + U_{\sigma\pi}|} 
\simeq  U_{\sigma\pi} - \frac{J_{H}}{4},
\end{equation}
since $\frac{|V_{\sigma \pi}|^2}{|\epsilon_d|}\sim 1$ meV is much smaller than $U_{\sigma \pi} \sim J_{H} \sim 100$ meV, according to the
estimates made in Sect.~\ref{sec:twolevel}. However, since  $U_{\sigma \pi} \sim J_{H}$ we cannot decide whether 
the triplet or the doublet are favored as the ground state of the dangling $\sigma$-$\pi$ zero-mode system. Thus, 
in the following we shall analyze both possibilities. 

\section{Schrieffer-Wolf-like projection}\label{sec:swt}

In order to integrate out the charge fluctuations, we shall project the Anderson model introduced above 
onto the subspace where the vacancy  dangling $\sigma$-level and the $\pi$ zero-mode are 
either in the triplet or doublet state. This requires that 
$\langle n_d \rangle \simeq 1$ and that $\langle n_0\rangle \simeq 1$ for the triplet case and $\langle n_0\rangle = 0$ for the doublet
case (see Sec.~\ref{sec:atomlimit} for a discussion of the validity of these approximations). 
To this end, we introduce a projection operator, $P$, 
 onto the subspace where $n_d = 1$ and either $n_0 = 1$ (triplet) or $n_0 = 0$ (doublet); $Q = 1 - P$  projects states onto the orthogonal subspace where either $n_d = 0$ or 
$n_d = 2$. The effective  Hamiltonian projected onto the $P$-invariant subspace can be obtained from the expression:
\begin{equation}
H_{\mathrm{eff}}(E) = PHP + PHQ \frac{1}{E-QHQ} QHP, \label{eq:heff}
\end{equation}
 which we can use to perturbatively powers of $V$ integrate out the fluctuations in the occupancy of the dangling $\sigma$  
level. However, the results, depend on whether the  ground state of the dangling $\sigma$ level $\pi$ zero-mode complex 
is a doublet or a triplet. 
\subsection{Doublet case}\label{subsec:swdoublet}
Let us assume that the ground state of the the dangling $\sigma$ level $\pi$ zero-mode complex is a doublet (i.e.
 $U_{\sigma\pi} > J_H/2$).  In that case $P$ is the projection operator onto the doublet state of the vacancy level system.
 To carry out the calculation, we first note that 
\begin{align}
PHP &= H_c  + P H_{\sigma\pi} P + PH_{\mathrm{hyb}}P, \\
QHQ &= H_c  + Q H_{\sigma\pi} Q + QH_{\mathrm{hyb}}Q,\\
PHQ &= PH_{\mathrm{hyp}}Q + PH_{\sigma\pi}Q. 
\end{align}
Upon neglecting retardation effects (i.e. setting $E = \mathcal{E}_0 + E_0(n_T = 1, S_T = \frac{1}{2}) \simeq \mathcal{E}_0 + \epsilon_d$, where  $\mathcal{E}_0$ is the ground state of the chain)  as well constant terms and using the identity \eqref{eq:sigmaiden}, 
we arrive at
\begin{align}
H_{K} &=  H_c   
+\frac{|V_0|^2 }{\epsilon_d}   P d^{\dag}_{\alpha} c^{\alpha}_{0} c^{\dag}_{0\beta} d^{\beta} P  
 - \frac{|V_0|^2}{U +\epsilon_d}  P c^{\dag}_{0\alpha} d^{\alpha} d^{\dag}_{\beta} c^{\beta}_0 P  \\ 
&= H_c      + J_K \mathbf{S}_d\cdot \mathbf{s}_0  
 + V_0 c^{\dag}_{0\alpha} c^{\alpha}_{0} + \frac{|V_0|^2 }{\epsilon_d},\label{eq:sdd}
\end{align}
where, to leading order in $V_0$, 
\begin{align}
J_K &= 2 |V_0|^2 \left[ -\frac{1}{\epsilon_d} + \frac{1}{U+\epsilon_d} \right] = 2 |V_0|^2 \left[ \frac{1}{|\epsilon_d|} + \frac{1}{U+\epsilon_d} \right]  > 0,\\
V_K &= - \frac{1}{2} |V_0|^2  \left[ \frac{1}{\epsilon_d}  + \frac{1}{U+\epsilon_d} \right] =   \frac{1}{2} |V_0|^2  \left[ \frac{1}{|\epsilon_d|}  - \frac{1}{U+\epsilon_d} \right] > 0
\end{align}
and 
\begin{align}
\mathbf{S}_{d} &= Pd^{\dag}_{\alpha} \left(\frac{\boldsymbol{\sigma}}{2}  \right)^{\alpha}_{\beta}  d^{\beta} P,\\
\mathbf{s}_0 &= c^{\dag}_{\alpha 0}    \left(\frac{\boldsymbol{\sigma}}{2}  \right)^{\alpha}_{\beta}  c^{\beta}_0
\end{align}

 \subsection{Triplet case}\label{subsec:swtriplet}
  
   We can repeat the same calculation as above by assuming that the ground state of the vacancy is a triplet 
state. In this case, $P$ projects onto the triplet state of the vacancy dangling $\sigma$ and $\pi$ zero-mode levels. 
Thus, neglecting retardation $E = \mathcal{E}_0 + E_0(n_T = 2, S_T = 1) = \mathcal{E}_0 + \epsilon_d + \varepsilon_0 + U_{\sigma\pi}- \frac{J_H}{4}$. Acting upon with the operator $H_{\mathrm{hyb}}$ then produces two kinds of excited states where  either $n_d = 0$ (having energy
$\simeq \mathcal{E}_0 + \varepsilon_0$ to leading order in $V$) or $n_d = 2$ (having energy $\simeq \mathcal{E}_0 + 2\epsilon_d + U + \varepsilon_0 + 2 U_{\sigma\pi}$ to leading order in $V$). Thus,
\begin{equation}
H_K =  H_c   - J_H P_t \mathbf{S}_d \cdot \mathbf{S}_0 P_t  + 2J_K P_t \mathbf{S}_d\cdot \mathbf{s}_0 P_t 
 + V_0 c^{\dag}_{0\alpha} c^{\alpha}_{0} \label{eq:hkt}
\end{equation}
In the above expression, $P_t$ is the projection operator onto the triplet state.
This results holds for temperatures below the double-triplet energy splitting $\simeq \frac{J_H}{4}- U_{\sigma\pi}$.
 As a consequence of rotational invariance,
\begin{equation}
P_t \mathbf{S}_d P_t = \frac{1}{2} \mathbf{S}_T,
\end{equation}
where $\mathbf{S}_{T} = \mathbf{S}_{d} + \mathbf{S}_0$ and $\mathbf{S}^2_T = S_T (S_T +1)$ being $S_T = 1$.
Thus, we can write Eq.~\eqref{eq:hkt} as a $S=1$-impurity Kondo Hamiltonian:
\begin{equation}
H_K = H_c + J_K \mathbf{S}_T \cdot \mathbf{s}_0 + V_K c^{\dag}_{\alpha 0} c^{\alpha}_0, \label{eq:sdt}
\end{equation}
where (recall that $U_{\sigma\pi} - \frac{J_H}{4} < 0$ for the triplet to be the ground state):
\begin{align}
J_K &=  |V_0|^2 \left[ \frac{1}{|\epsilon_d + U_{\sigma\pi} - \frac{J_H}{4}|} + \frac{1}{U+\epsilon_d +U_{\sigma\pi} 
- \frac{J_H}{4}} \right] > 0,\\
V_0 &= \frac{1}{2} |V_0|^2  \left[ \frac{1}{|\epsilon_d + U_{\sigma\pi} - \frac{J_H}{4}|}  - \frac{1}{U+\epsilon_d + U_{\sigma\pi} 
- \frac{J_H}{4}} \right] > 0.
\end{align}
In the above derivation, we have used that $P_t \mathbf{S}_d \cdot \mathbf{S}_0P_t = \frac{1}{2}
 P_t  \left[ \mathbf{S}^2_T - \mathbf{S}^2_0 - \mathbf{S}^2_d \right] P_t = \frac{1}{4}$ and dropped the resulting
 constant term.

\section{Solution of the Scaling Equations for the Kondo coupling}

 Let us next consider the  solution of the scaling equations  for  Kondo coupling:
 \begin{equation}
\frac{d j_K}{d \ln D} =  \frac{d \ln \Delta(\epsilon)}{d\ln \epsilon} \Big|_{\epsilon=D} j_K - j_K^2. \label{eq:arg}
 \end{equation}
To find the solution to this equation, let us use the ansatz:
\begin{equation}
j_K(D) = f(D) g(D), \quad \frac{dj(D)}{d\ln D} = \frac{df(D)}{d\ln D} g(D)  + f(D) \frac{dg(D)}{d\ln D},
\end{equation}
and require that our choice of $f(D)$ cancels the first term on the right hand-side of \eqref{eq:arg}, which leads to
\begin{equation}
\frac{d}{d \ln D} \left( \frac{f(D)}{\Delta(\epsilon = D)}\right) = 0,
\end{equation}
and therefore $f(D) = C \Delta(\epsilon = D)$, where $C$ is a constant. Upon choosing $C = 1$, without loss of 
generality,  we find that $g(D)$ obeys the following equation:
\begin{equation}
\frac{d g(D)}{d \ln D} = -\Delta(\epsilon = D) g^2(D), 
\end{equation}
which can be integrated from $D = D_0$ to $D = T$ to  yield ($g(0) = g(D_0)$):
\begin{equation}
\frac{1}{g(T)} - \frac{1}{g(0)} = \int^{T}_{D_0} \frac{\Delta(\epsilon)}{\epsilon} 
\end{equation}
Next we recall that $j(T) = \Delta(\epsilon = T) g(T)$ and therefore,
\begin{equation}
g(T) = \frac{j(T)}{\Delta(T)}.
\end{equation}
Hence,
\begin{equation}
\frac{\Delta(T)}{j(T)} - \frac{\Delta(D_0)}{j(0)} =  \int^{T}_{D_0} \frac{\Delta(\epsilon)}{\epsilon},
\end{equation}
which, when solved for $j(T)$ yields:
\begin{equation}
j(T) = \frac{\Delta(T) j(0)/\Delta(D_0)}{1 -  \frac{j(0)}{\Delta(D_0)}\int^{T}_{D_0} \frac{\Delta(\epsilon)}{\epsilon}, }
\end{equation}
Again, we can define the Kondo temperature as the value of $T$ where $j(T) = C^{-1}\sim 1$, which form the previous
expression leads to:
\begin{equation}
\int^{T_K}_{D_0} \frac{\Delta(\epsilon)}{\epsilon} - C\Delta(T_K)  = \frac{\Delta(D_0)}{j(0)} \propto \frac{1}{J_K}.
\end{equation}
Introducing $\Delta(\epsilon) \sim \left[ |\epsilon| \ln^2(\epsilon/D_0) \right]$~\cite{Pereira2006Nunob} in the above equation leads to 
to the condition that $T_K \ln^2 (T_K/D_0) \sim J_K$ quoted in the main text. 

\section{Slave-boson Mean-field Solution}

In the present study we employ the slave bosons technique combined with the large -$N$ mean field treatment of the resulting action. The slave bosons technique aims at enlarging the Hilbert space of the impurity by introducing a slave boson creation operator $b^\dagger$ and a new fermion $f^\dagger_{\sigma}$  that act on a new vacuum state $\vert\textrm{vac}\rangle$ such that the physical empty and singly occupied physical states of the impurity orbital are represented as
\begin{align}
\vert0\rangle&=b^{\dagger}\vert\text{vac}\rangle,\\
\vert\sigma\rangle&=f_{\sigma}^{\dagger}\vert\text{vac}\rangle.
\end{align}
In terms of the slave boson and the new fermion the original impurity creation operator reads $d_{\sigma}^{\dagger}=f_{\sigma}^{\dagger}b$, which satisfies canonical anti-commutation relations provided the Hilbert space is restricted to the physical space. The spin degeneracy is generalized from SU(2) to SU($N$), with $\sigma=1,\ldots N$, and we shall consider the $U=\infty$ limit with a constraint of no double occupancy (for $N=2$) generalized to
\begin{equation}
b^{\dagger}b+\sum_{\sigma}f_{\sigma}^{\dagger}f_{\sigma}=\frac{N}{2}
\end{equation}
for arbitrary $N$. The resulting effective action is:
\begin{multline}
S_{\mathrm{eff}}=\int_{0}^{\beta}d\tau\left\{ \sum_{\sigma}f_{\sigma}^{\dagger}(\partial_{\tau}+\varepsilon_{d})f_{\sigma}+b^{\dagger}\partial_{\tau}b  + i\lambda(\tau)\left[b^{\dagger}b+\sum_{\sigma}f_{\sigma}^{\dagger}f_{\sigma}-\frac{N}{2}\right]\right\} \\
-\frac{2|V|^2}{N}\int_{0}^{\beta}d\tau\int_{0}^{\beta}d\tau'\sum_{\sigma}f_{\sigma}^{\dagger}(\tau)b(\tau)g(\tau-\tau')b^{\dagger}(\tau')f_{\sigma}(\tau')
\end{multline}
where the Lagrange multiplier $\lambda$ is a dynamical variable introduced to impose the constraint.

In the large $N$ limit, the partition function can be approximated by means of a saddle point evaluation of its defining functional integral. At the saddle point, the Lagrange multiplier becomes static $i\lambda(\tau)\to\langle i\lambda\rangle = \lambda_0$ and the boson field is replace by its average $b\to \langle b\rangle = \sqrt{N/2} r_0$ and $b^\dagger\to\langle b^\dagger \rangle = \sqrt{N/2}r_0$. The values of $r_0$ and $\lambda_0$ at the saddle point are determined by minimizing of the free energy, which leads to the following mean field equations
\begin{align}
\frac{r_{0}^{2}}{2}+\frac{1}{\beta}\sum_{\omega_{n}}G_{f}(i\omega_{n})e^{i\omega_{n}0^{+}}&=\frac{1}{2}\label{eq:1_slave_boson_mean_field}\\
\lambda_{0}+\frac{2|V|^2}{\beta}\sum_{\omega_{n}}G_{f}(i\omega_{n})g(i\omega_{n})e^{i\omega_{n}0^{+}}&=0\label{eq:2_slave_boson_mean_field},
\end{align}
where $G_f$ is the fermion propagator, which reads:
\begin{equation}
G_{f}(i\omega_{n})=\langle f_{\sigma}^{\dagger}(i\omega_{n})f_{\sigma}(i\omega_{n})\rangle  =\frac{1}{i\omega_{n}-\varepsilon_{d}-\lambda_{0}-r_{0}^{2}V^2g(i\omega_{n})}.
\end{equation}
We fix the Kondo temperature $T_K$ at the onset of condensation of the slave boson, $r_0=0$. $G_{f}$ reduces to the standard Green's function for a free fermion, and therefore Eq. (\ref{eq:1_slave_boson_mean_field}) leads to 
\begin{equation}
f_{F}\left(\frac{\varepsilon_{d}+\lambda}{T}\right)=\frac{1}{2}
\end{equation}
where $f_{F}(u)=1/(e^{u}+1)$ is the Fermi factor. Its solution is given by $\lambda_0=-\varepsilon_{d}$. Thus,  Eq. (\ref{eq:2_slave_boson_mean_field}) becomes:
\begin{equation}
\frac{2}{\beta}\sum_{\omega_{n}}\frac{1}{i\omega_{n}}g(i\omega_{n})e^{i\omega_{n}0^{+}}=\frac{\varepsilon_{d}}{|V|^{2}},
\end{equation}
which we have solved numerically by first computing the local Green's function a the tight-binding model of the graphene
including nearest neigbor $t$ and next-nearest neighbor hopping $t^{\prime}$.

\end{document}